 \documentclass[nohyper,12pt,letterpaper]{JHEP3}
 \usepackage{epsfig}
 
 \title{Defect CFTs and holographic multiverse}
 \author{Bartomeu  Fiol\\
 Departament de F{\'\i}sica Fonamental i \\Institut de Ci{\`e}ncies del Cosmos, 

Universitat de Barcelona,

Mart{\'\i}\ i Franqu{\`e}s 1, 08193 Barcelona, Spain.\\

\email{bfiol@ub.edu}}

\abstract{We investigate some aspects of a recent proposal for a holographic description of the multiverse. Specifically, we focus on the implications on the suggested duality of the fluctuations of a bubble separating two universes with different cosmological constants. We do so by considering a 
similar problem in a 2+1 CFT with a codimension one defect, obtained by an M5-brane probe embedding in $AdS_4\times S^7$, and studying its spectrum of fluctuations. Our results suggest that the kind of behavior required by the spectrum of bubble fluctuations is not likely to take place in defect CFTs with an AdS dual, although it might be possible if the defect supports a non-unitary theory.}

\begin{document}

\section{Introduction}
In the quest to formulate a quantum theory of gravity, the holographic principle \cite{'tHooft:1993gx} is an appealing model-independent guiding principle, that has received a tremendous impetus after finding a concrete realization in the AdS/CFT correspondence \cite{Maldacena:1997re}. 

Given the wide and deep ramifications of the AdS/CFT correspondence, it is natural to try to realize the holographic principle as a way to formulate quantum gravity in spacetimes with different asymptotia, although there is no a priori guarantee that the problem is well defined or has a manageable answer.
Since inflation plays a central role in modern cosmology, it has attracted a fair share of activity trying to provide a holographic dual for various inflationary models. One of the first ideas in this direction is due to Strominger, who suggested \cite{Strominger:2001gp} that a single inflating bubble is dual to a renormalization group flow between two Euclidean conformal fixed points. In this picture, the theory at the future infinity boundary would be the UV fixed point theory, so the time evolution of the universe would be dual to an upstream renormalization group flow. There have been many works trying to extend this proposal to the scenario of eternal inflation \cite{Freivogel:2006xu, Freivogel:2009rf, Bousso:2009dm}, out of which we will focusing our attention in the recent proposal of the holographic multiverse \cite{Garriga:2008ks, Garriga:2009hy}.

The holographic multiverse proposal claims that there exists a holographic dual to eternal inflation, and that, as in \cite{Strominger:2001gp}, it can be defined at the future infinity boundary of spacetime. Since now future infinity is boundary to regions in spacetime with different cosmological constants, it is expected that the number of degrees of freedom is not constant along the 3d boundary, but just constant on patches, with the particular value of a domain in the boundary tied to the Hubble expansion parameter of the corresponding region in the bulk of spacetime. Furthermore, the authors of \cite{Garriga:2008ks, Garriga:2009hy} left open the possibility that additional degrees of freedom live at the 2d walls separating 3d domains of the boundary.

In this note we want to focus precisely on these 2d walls on the hypothetical dual theory, and the implications that the dynamics of bubbles in the bulk of spacetime have for them. As we will argue, these issues lead naturally to consider defect conformal field theories, i.e. conformal field theories with a defect of lower dimension that preserves a subgroup of the original conformal symmetry group \cite{Cardy:1984bb}. Defect conformal field theories have been studied using the AdS/CFT correspondence \cite{Karch:2000gx, DeWolfe:2001pq, Aharony:2003qf,  Constable:2002xt}, and we will also rely on this correspondence for our specific computations.

To set the stage, consider the simplest scenario: two 4d deSitter solutions with different values of the cosmological constant, separated by a thin wall bubble. In the hypothetical dual theory defined at the future infinity boundary, this ought to correspond to two 3d regions with different UV CFTs (and different number of degrees of freedom), separated by a codimension one defect with some additional degrees of freedom living on it. As we review below, one can study the fluctuation modes of the bubble to try to narrow down the possible dual theory. The analysis of the bubble fluctuations is standard, but trying to interpret them as coming from a dual theory leads to some counterintuitive results \cite{Garriga:2009hy}; for instance, it was found that the quantity expected to correspond to the number of degrees of freedom with non-trivial boundary conditions at the wall can actually be smaller than the difference between the number of degrees of freedom on the two sides. This puzzling result implies that if the proposed duality does exist, we must sharpen our understanding on how it maps bubbles to defects in the dual theory, and what kind of interactions it requires among the defect degrees of freedom and the ambient ones in the dual theory. The purpose of this note is to improve our very limited casuistics, by working out a concrete example with a known 2+1 CFT.

In setting up a field theory computation, we have to choose the three dimensional field theory and the two dimensional defect we are going to add to it. Let's motivate our choices in turns. As for the choice of field theory, a possibility might be to start with a free field theory, but one should have reservations about a free field theory correctly capturing a regime where gravity is weakly coupled\footnote{On the other hand, 3d super-renormalizable theories might be candidates for a holographic description of inflation in a regime where gravity is strongly coupled \cite{McFadden:2010na}.}; we will not pursue this direction. If on the other hand, we consider a strongly coupled 3d CFT, our chances of performing a computation directly in the CFT are severely limited. As a way out of this conundrum, I will consider a 3d CFT that has a known AdS dual, namely, the IR limit of maximally supersymmetric 2+1 $SU(N)$ SYM theory, which is expected to be dual to M-theory on $AdS_4\times S^7$.  In the large N limit, this reduces to something we can work with, 11D SUGRA on $AdS_4\times S^7$; having to take the large N limit is presumably not a limitation for the problem at hand, since we are interested in understanding the putative dual of large space-times, which should involve some sort of large N limit. 

A potential reservation is that while our aim is trying to make sense of some puzzling properties of a hypothetical Euclidean theory, which is presumably non-unitary \cite{Strominger:2001pn}, we consider a Lorentzian unitary 2+1 field theory. While it is clear that the field theory we consider has to differ in some significant way from the theory we are after, the reason that our choice is eventually useful is that the computations we perform using AdS/CFT end up tracking  closely the ones performed in \cite{Garriga:2009hy}, so we can understand the different outcomes easily.

As for the defect, we want a 1+1 wall in the 2+1 boundary. We can accomplish that by suitably embedding a probe brane in the background. Furthermore we want that the amount of flux on both sides of the brane is different, reflecting a difference in the value of the vacuum density energies. We accomplish this by adding a probe M5-brane with world-volume magnetic flux turned on. The M5-brane world-volume is $AdS_3\times S^3$, with the $S^3$ piece inside $S^7$, so from the four dimensional point of view it is a domain wall.


We consider fluctuations about this probe brane embedding, and extract the contribution to the trace anomaly due to the 1+1 wall in the boundary. The results that we find for this specific example do not display any of the counterintuitive features encountered in \cite{Garriga:2009hy}. In the case at hand, the key difference appears to be that while in the $dS$ computations the curvature radius of the bubble is always equal or smaller than the Hubble expansion parameter, in $AdS$ the opposite is true. Although we have worked out a single example, this particular feature is common for $AdS_p$ branes in $AdS_q$ spaces, and we believe that this feature is the reflection on the gravity side that the behavior of the boundary degrees of freedom conforms to intuitive expectations. 

Given these results, there are various attitudes one can take. The first possibility is to take them as evidence that the duality proposed in \cite{Garriga:2008ks, Garriga:2009hy} does not make sense. A less drastic possibility is that the type of unusual behavior encountered in \cite{Garriga:2009hy}  doesn't happen for Lorentzian CFTs which admit an AdS dual; it should be kept in mind that the Euclidean 3d CFT conjectured in \cite{Garriga:2008ks, Garriga:2009hy} does not necessarily correspond to the Wick rotation of a sensible Lorentzian CFT. In the discussion section, we suggest that if the holographic multiverse proposal makes sense, the degrees of freedom in the defect might have to be non-unitarity.

The plan of the paper is as follows. In section 2, we review the results of \cite{Garriga:2009hy} for bubble fluctuations and the difficulties they pose in trying to interpret them holographically. In section 3,  we present the setup of the computation we are going to perform, by introducing the supergravity background dual to the CFT we consider, and the M5 probe brane that introduces the defect on the CFT. In section 4 we focus on a particular fluctuation mode around the solution, and compute its regularized action, which allows us to read off the contribution of this defect to the integrated trace anomaly of the CFT, and compare the result with those found in \cite{Garriga:2009hy}. We conclude with some discussion about possible implications for the hypothesized duality of \cite{Garriga:2008ks, Garriga:2009hy}.

\section{Bubble fluctuations}
The authors of \cite{Garriga:2009hy} proposed that eternal inflation is holographically dual to an Euclidean 3-dimensional field theory, controlled in the UV by a conformal fixed point. Their proposal extends ideas of Strominger about inflation as an upstream renormalization group flow \cite{Strominger:2001gp}. 

This CFT, and the renormalization group flow that describes the multiverse, if they exist at all, are probably bewilderingly complicated.  As a way to extract concrete predictions out of their general proposal, the authors of \cite{Garriga:2009hy}  considered a simplified scenario where a single spherical bubble separates two universes; the bubble has a thin wall with $dS_3$ metric separating two $dS_4$ spaces, that in general have different vacuum density energy. Analysis of fluctuations of this type of bubbles have been around for a while \cite{Garriga:1991ts}; the novel question addressed in \cite{Garriga:2009hy} is what can one deduce about the potential holographic dual from the knowledge of the bubble fluctuation spectrum.  In this section we review their results, recasting them in a language more common in the AdS/CFT literature.

The bulk spacetime is taken to be $dS_4$. It is best to consider $dS_4$ in flat coordinates
$$
ds^2_{dS_4}=-dt^2+e^{2Ht}\left(dr^2+r^2d\Omega_2\right)
$$ 
since it makes manifest the identification of time evolution in the bulk with scale transformations in the boundary \cite{Strominger:2001gp}, while for the  induced metric on the $dS_3$ bubble we use conformal coordinates
$$
ds^2_{dS_3}=\left(\frac{R_0}{\cos \eta}\right)^2\left(-d\eta^2+d\Omega_2\right)\equiv \tilde a(\eta)^2
\left(-d\eta^2+d\Omega_2\right)
$$
where $R_0$ is the curvature radius of the bubble. In the thin wall approximation one can express $R_0$ in terms of the tension $T$ of the bubble, the Hubble expansion parameter $H$, and the jump in the vacuum energy density between the interior and the exterior of the bubble $\Delta \rho_V$ \cite{Berezin:1987bc},
\begin{equation}
R_0^2 \sim \frac{9T^2}{9H^2T^2+(\Delta \rho_V)^2}
\label{thinwall}
\end{equation}
As expected on geometric grounds, it follows from this formula that the curvature radius of the wall is always equal or smaller than that of the background, $R_0\leq H^{-1}$.

In the thin wall approximation, the only fluctuations considered are the ones associated with the transverse position of the bubble wall. These can be captured by a scalar field $\phi$ living on the wall \cite{Garriga:1991ts}. For the specific case at hand, a thin $dS_3$ wall in $dS_4$, it turns out that this scalar field is tachyonic, with mass \cite{Garriga:1991ts}
\begin{equation}
m^2_\phi=-\frac{3}{R_0^2}
\label{tacmass}
\end{equation}
This tachyonic mass can be understood as indicating that as the bubble expands, the fluctuations grow accordingly.

In trying to pin down the possible dual field theory, the first thing that might come to mind is to determine the central charge of the UV CFT in terms of gravitational data, following the steps of \cite{Brown:1986nw, Strominger:2001pn}. At first, this idea doesn't look promising, as  3d CFTs defined on manifolds without boundaries have no trace anomaly.  However, the presence of a domain wall (a 2d defect) can change that, if the CFT fields are required to satisfy non-trivial boundary conditions on it. A way to think of this is to consider that there are additional degrees of freedom living on the defect (placed at, say, $x=0$), so the full stress-energy tensor of the boundary theory is \cite{DeWolfe:2001pq, Aharony:2003qf}
$$
T_{\mu \nu}^{TOTAL}=T_{\mu \nu}^{3d}+\delta(x)T^{2d}_{ij}\delta ^i_\mu \delta^j_\nu
$$
and the defect stress-energy tensor $T^{2d}_{ij}$ can give rise to a non-zero integrated trace anomaly, even if $T^{3d}_{\mu \nu}$ cannot.
Indeed, if $K_{ij}$ and $\hat R_{ij}$ are the extrinsic and intrinsic curvatures of the 2d defect, the integrated trace anomaly is given by an expression of the form \cite{Schwimmer:2008yh}
\begin{equation}
a_{3/2}=\int d \Sigma_2 \sqrt{\hat g}\left[d_1(K_{ab}K^{ab}-\frac{1}{2}K^2)+d_2\hat R\right] 
\label{confan}
\end{equation}
where we already assumed that the ambient 3d space is flat. The coefficients $d_1$ and $d_2$ depend on the specific theory. The term with $\hat R$ is the usual
topological term in 2d, while the first term is only possible because the 2d defect is embedded into a 3d ambient space. This first term is sensitive to variations of the shape of the defect, and it was argued in \cite{Garriga:2009hy} that it is holographically dual to the action that governs the fluctuations of the bubble. This is in accordance with the standard AdS/dCFT dictionary \cite{Karch:2000gx, DeWolfe:2001pq}, where the probe brane degrees of freedom (bubble degrees of freedom in our case) are holographically dual to degrees of freedom on the defect (the 2d wall in our case).

The arguments of \cite{Garriga:2009hy} allow to estimate the coefficient $d_1$ of the dCFT, through a computation of the action for the fluctuations of the bubble. A sketch of their computation goes as follows: while for an observer on the unperturbed brane the field that naturally captures fluctuations of the brane is the one measuring deviations normal to the brane - the field $\phi$ mentioned above - to make contact with a possible boundary theory, it is more convenient to trade it by a field, dubbed  $\delta$ in \cite{Garriga:2009hy}, that measures fluctuations for slices of constant $t$. An straightfoward evaluation of the action for these fluctuations yields a divergent result, as it is well-known to happen in AdS/CFT for the bulk holographic action, both for the full CFT and for submanifold observables. It is therefore necessary to introduce a cut-off value for the scale factor $\tilde a(\eta)$ and consider a regularized holographic action. It turns out that this regularized action contains a term logarithmic in the $\tilde a(\eta)$ cut-off , just as it is the case for even dimensional CFTs \cite{Henningson:1998ey} and even dimensional submanifold observables \cite{Berenstein:1998ij, Graham:1999pm}. This logarithmic term induces a conformal anomaly for the finite piece of the regularized action. Thus, the anomaly of the renormalized action is fixed in terms of the coefficient of the logarithm term in the regularized action. In the case at hand, it is argued in \cite{Garriga:2009hy} that the log term of the regularized bubble action corresponds to the term with coefficient $d_1$ in the expression for the integrated trace anomaly (\ref{confan}), allowing to determine it. Omitting numerical factors, this is given by
\begin{equation}
d_1\sim \frac{TR_0}{H^2}
\label{resultd}
\end{equation}
The authors of \cite{Garriga:2009hy} further interpreted $d_1$ as roughly a measure of the number of degrees of freedom of the hypothetical dual field theory that must satisfy non-trivial boundary conditions at the defect (their reasoning being that if the wall was transparent to all the degrees of freedom, there would be no anomaly).  However, this interpretation of $d_1$ and the estimate (\ref{resultd}) are somewhat at odds. A first conterintuitive feature of this result is that for fixed $T$ and $H$, as $\Delta \rho_V$ increases, it follows from (\ref{thinwall}) that $R_0$ decreases, and so does $d_1$. This would imply that as the two universes separated by the wall differ more, there are {\it less} degrees of freedom satisfying non-trivial boundary conditions at the wall. 

A second counterintuitive feature, emphasized already in \cite{Garriga:2009hy} is that, if we denote by $c_i$ the number of degrees of freedom on each side of the wall, $\Delta c$ should be a solid lower bound on the number of degrees of freedom with non-trivial boundary conditions, as it gives the difference of degrees on freedom between the two sides. Therefore, this interpretation of $d_1$ leads to the conclusion that
\begin{equation}
d_1\geq \Delta c
\label{puzzle}
\end{equation}
However, it can be shown \cite{Garriga:2009hy} that this implies, within the thin wall approximation, that $R_0\sim H^{-1}$. One arrives then at the conclusion that the regime $R_0\ll H^{-1}$, while it can be reliably considered within the thin wall approximation, leads to a violation of (\ref{puzzle}), so the 
intuition that led to the previous inequality must fail in this regime.

To clarify the implications of these results, it seems appropriate to sharpen our intuition about the meaning and parametric dependence of $d_1$ in a situation when we do have a CFT realization. We turn now to that task.

\section{An $AdS_4$/dCFT computation: the setup}
The 2+1 CFT we are going to consider is the IR fixed point of maximally supersymmetric 2+1 $SU(N)$ SYM. An explicit Lagrangian description of this CFT became available only recently \cite{Aharony:2008ug}, in terms of a supersymmetric Chern-Simons theory with two gauge groups coupled to matter. 
This conformal theory does not have a marginal coupling that allows us to go to a free field theory limit: the theory is inherently strongly coupled, so direct computations on the CFT would be quite hard to perform. Furthermore, while there has been some work in providing  a direct description of defects on this CFT, the results are not conclusive \cite{Ammon:2009wc}. Since we are not going to be using this Lagrangian description, we won't present it here, referring the interested reader to \cite{Aharony:2008ug} for further details.

For our purposes, what is important about this CFT is that it is the low energy limit of the world-volume theory of a stack of N M2-branes, and it is conjectured to be dual to M-theory on $AdS_4\times S^7$. In the large $N$ limit, this reduces to 11d SUGRA on the same background, becoming a tractable setup that allows explicit computations. Furthermore, it is straightforward to add a probe brane that reaches the boundary at a codimension one defect. To make contact with the discussion of the previous section, we are interested in having a defect that separates two CFTs with different numbers of degrees of freedom, and for that we will use the general idea of Karch and Randall \cite{Karch:2000gx} of having some of the M2-branes ending on the M5-brane probe.  This means that the probe brane presents a bending, compensated by a non-trivial magnetic flux on the world-volume. A difference with the discussion in the previous section is that the defect we will consider here is a flat 1+1 wall, rather than a sphere. 

Before we embark in the actual computations, rethinking the arguments of  \cite{Garriga:2009hy} in the light of AdS/dCFT leads to a refined interpretation of the coefficient $d_1$. When one considers defect CFTs in the AdS/CFT correspondence by adding a probe brane to an AdS background, there are two kinds of {\it new} degrees of freedom,  those living on the brane and those living on the defect, and
it is expected that holography acts a 'second time' making them dual to each other in the appropriate limit. For instance, in the extensively studied case of a D5 probe in the background of D3 branes  \cite{Karch:2000gx, DeWolfe:2001pq}, there are open strings with both ends on the D5 probe, that are additional degrees of freedom on the gravity side, and in the decoupling limit they are expected to be holographically dual to the degrees of freedom coming from open strings coming from 3-5 strings, that live on the defect, and enlarge the content of the boundary field theory. Since we are using brane fluctuations to determine the coefficient $d_1$, this coefficient must be sensitive to these {\it new} degrees of freedom on the defect. This interpretation of $d_1$ does not neccesarily clash with the previous one: in fact, ordinarily the degrees of freedom on the defect can be seen as enforcers of specific boundary conditions for the ambient fields at the defect, and can even be integrated out, leaving its mark as non-trivial boundary conditions for the ambient fields (see \cite{Gomis:2006sb} for a nice illustration of this). However, given the subtleties of the present problem,  we prefer to keep these two points of view in sight.

This interpretation already eases the tension in comparing $d_1$ with $\Delta c$, since $d_1$ is not immediately counting ambient degrees of freedom already present before adding the defect, and details about the couplings between defect and ambient fields in the boundary theory might alter the arguments leading to the inequality (\ref{puzzle}). Nevertheless, it still leaves open the interpretation of the fact that in deSitter, $d_1$ decreases as the difference in vacuum energy on the two sides of the wall increases. We will return to this point in the discussion section.

\subsection{The background}
The bosonic content of 11d SUGRA is the 11d metric and a 3-form $C_{3}$. A maximally supersymmetric solution has metric of the form $AdS_4 (R/2) \times S^7(R)$ for any value of $R$.
We write the $AdS_4\times S^7$ solution in a way that makes manifest a 2+1 Minkowski boundary,
$$
ds^2=\frac{r^4}{R^4}dx_{1,2}^2+\frac{R^2}{r^2}d\vec r^2 
$$
$$
 C_{(3)}=\frac{r^6}{R^6}dx^0\wedge dx^1\wedge dx^2
$$
where $\vec r$ is an 8-dimensional vector in $R^8$. For what follows, it is convenient to split this 8d vector into two 4-d vectors, $\vec y$ and $\vec z$, with $\rho$ the norm of $\vec y$. The background metric reads then
\begin{equation}
ds^2=\frac{(\rho^2+\vec z^2)^2}{R^4}dx^2_{1,2}+\frac{R^2}{\rho^2+\vec z^2}\left(d\rho^2+\rho^2 d\Omega_3^2+d\vec z^2\right)
\label{pickst}
\end{equation}
The reason to perform this split is to make manifest a choice of an $S^3$ inside $S^7$, that will become handy next, when we discuss the embedding of the probe brane.

\subsection{Adding a defect: the probe brane}
The supergravity solution just presented is expected to be dual to a Lorentz invariant 2+1 CFT. We now want to add a codimension one defect on the CFT.  These defects haven't been much studied directly for the CFT at hand, but we can work around that by considering the dual description of adding a defect to the CFT. This is given by considering a probe brane in the previous geometry, such that it reaches the boundary of $AdS_4$ as a codimension one defect.

The probe we consider is an M5-brane. The world-volume of a single M5-brane consists of a 6d $(2,0)$ tensor multiplet, whose bosonic content is given by five scalar fields and a self-dual three-form field strenght $F=dB$. The fact that the three-form on the world-volume of the M5-brane is self-dual makes it notoriously difficult to write down an action. A way to circumvect this problem is to introduce an extra non-dynamical scalar field $a(\xi)$, with a non-polynomial action; we will follow this route and as the world-volume action of the M5 brane we will take the PST action \cite{Pasti:1997gx},
\begin{equation}
S_{PST}=T_{M5}\int d^6\xi \left[-\sqrt{-|g_{ij}+\check H_{ij}|}+\frac{\sqrt{-|g|}}{4\partial a \cdot \partial a}
\partial _i a (*H)^{ijk}H_{jkl}\partial^l a\right] +T_{M5}\int \frac{1}{2}F\wedge P[C_{(3)}]
\label{pstact}
\end{equation}
where $H$ is a 3-form that combines $F$ and the pullback on the world-volume of the background 3-form potential $C_{(3)}$,
$$
H=F-P[C_{(3)}]
$$
and
$$
\check H^{ij}=\frac{1}{3!\sqrt{-|g|}}\frac{1}{\sqrt{-(\partial a)^2}}\epsilon^{ijklmn}\partial _k a H_{lmn}
$$
Now we look for a M5 embedding in the $AdS_4\times S^7$ geometry. Since we want a codimension one defect on the 2+1 boundary, we take $x^0,x^1$ as two of the world-volume coordinates. We also take $\rho$ as a world-volume coordinate, so the brane expands in the $AdS_4$ bulk as well. Finally, we identify the three remaining world-volume coordinates with the coordinates of the $S^3$ inside $S^7$ that we singled out in (\ref{pickst}).

With this identification of world-volume coordinates, a simple solution for the embedding equations has $x^2$ constant and the rest of the world-volume fields set to zero. This solution has world-volume metric $AdS_3\times S^3$ and reaches the boundary at a codimension one defect, but for our purposes it has the fatal drawback that the bulk on the two sides is the same. A way to get a jump in the number of degrees of freedom between the two sides of the brane was discussed by Karch and Randall \cite{Karch:2000gx}. In the brane configuration, before taking the near-horizon limit, it consists in having $k$ M2-branes ending on the M5. This causes a non-trivial bending on the M5, implying we must allow for a non-trivial $x^2(\rho)$, which is accompanied by a magnetic flux for three-form living on the world-volume on the brane, in the internal $S^3$. Armed with this insight, as ansatz for the remaining world-volume fields we take 
$$
x^2=x^2(\rho),\hspace{1cm}\vec z=0, \hspace{1cm} F=q \hbox{ vol} (S^3)
$$
and following \cite{Arean:2007nh}
fix the gauge $a=x^1$. This ansatz is actually a particular case of the one considered in \cite{Arean:2007nh}. They consider a more general ansatz, with $|\vec z|=L$, that gives mass to some fields of the dual CFT, breaking conformal invariance. This more general ansatz might be relevant if we want to use this type of constructions to study the proposed duality beyond its UV fixed point, but in this work we will focus on this fixed point limit, and  keep $\vec z=0$. The solution found in \cite{Pasti:1997gx, Arean:2007nh} for this ansatz is
\begin{equation}
x^2(\rho)=x^2(\infty)+\frac{q}{2\rho^2}
\label{solprobe}
\end{equation}
It can be checked that this is a $1/2$ BPS solution. The induced metric on the M5-brane world-volume is
\begin{equation}
ds^2_{M5}=\frac{\rho^4}{R^4}dx^2_{1,1}+R^2 \left(1+\frac{q^2}{R^6}\right) \frac{d\rho^2}{\rho^2}+R^2d\Omega_3^2
\label{wvmetric}
\end{equation}
so it is of the form $AdS_3(R_{eff}/2)\times S^3(R)$, with
\begin{equation}
R_{eff}^2=R^2\left(1+\frac{q^2}{R^6}\right)
\label{radeff}
\end{equation}
We see that as consequence of a non-zero magnetic flux ($q\neq 0$), the curvature radius of the wall $R_{eff}$ is no longer equal to the Anti de Sitter radius $R$, but contrary to what happened for the thin wall in de Sitter, it's always bigger, $R_{eff}\geq R$. This obvious difference between de Sitter and anti de Sitter embeddings appears to be at the root of the difficulties in giving an intuitive interpretation of (\ref{resultd}).

\section{Computing the conformal anomaly}
Our next task is to compute the coefficient $d_1$ in the conformal anomaly (\ref{confan}),  for the defect we have presented in the previous section. As reviewed in section 2, in \cite{Garriga:2009hy} it was pointed out that a way to do so is to consider fluctuations of the brane inducing the defect, consider a regularized action with a cut-off in the holographic direction, and read off the coefficient from the term with the logarithmic dependence on the cut-off.

\subsection{Fluctuations}
To compute the renormalized action for fluctuation of the position of the brane, we need to know the Lagrangian controlling the fluctuations of the world-volume fields. This has been extensively studied in the case of D-branes, where the metric that appears is the open string metric, which needs not to coincide with the closed string metric (i.e. the induced world-volume metric). For M5-branes, it was argued in \cite{Gibbons:2000ck} that a close analog of the open string metric also controls the fluctuations, and this has been checked for a subset of the fluctuations \cite{Arean:2007nh}. A detailed analysis of the fluctuation spectrum, including fluctuations of all the world-volume fields, will be presented elsewhere \cite{inprep}. 

We are going to consider just fluctuations of the position of the brane, but since in general the equations of motion for the fluctuations are coupled, we need to argue that it is consistent to do so. First, it turns out \cite{inprep} that fluctuations of the scalars describing transverse directions to the $S^3$ in $S^7$ decouple from the rest, as it happens in analogous D-brane systems \cite{DeWolfe:2001pq}, so it is consistent to set them to zero. Second, since the fluctuation of the position of the brane $x^2$ is a scalar in the $AdS_3$ part of the M5 world-volume, in the equations of motion of the fluctuations of the 2-form that have some $S^3$ index, it only appears with derivatives with respect to $S^3$ coordinates. The upshot is that, as long as we consider an $x^2$ fluctuation independent of $S^3$ coordinates, it decouples from the rest, and it is consistent to consider it on its own. This suits us well, since in the previous section the fluctuation of the $dS_3$ brane didn't depend on any internal coordinates either. So we consider small fluctuations of $x^2$ around the solution (\ref{solprobe}),
$$
x^2(\xi)=x^2_{sol}+h(\xi)
$$
and expand the action (\ref{pstact}) around the solution up to quadratic order in $h(\xi)$. Before expanding the action, it is convenient to rewrite the second term in (\ref{pstact}) as
$$
\frac{\sqrt{-|g|}}{4\partial a \cdot \partial a}\partial _i a (*H)^{ijk}H_{jkl}\partial^l a=
\frac{1}{4!}\epsilon^{1jklmn}H_{1jk}H_{lmn}
$$
which upon inspection makes evident that the last two terms in (\ref{pstact}) contribute only a linear term in $h(\xi)$, coming from the pullback of $C_{(3)}$. This linear term cancels a similar linear contribution from the first term in (\ref{pstact}), and we conclude that all the quadratic fluctuations come from the square root in (\ref{pstact}). While carrying out the expansion is quite tedious, the final result is reassuringly simple. The Lagrangian that controls the fluctuation of the transverse position is
\begin{equation}
{\cal L}_{fluc}=-\frac{1}{2}\left(\rho^3\frac{R_{eff}^2}{R^2}\sqrt{|g_{S^3}|}\right)\frac{\rho^4}{R^2R^2_{eff}}
\tilde G^{ij}\partial _i h \partial _j h
\label{fluclag}
\end{equation}
The quantity in parentheses is the determinant of the square-root part of the action (\ref{pstact}), evaluated at the solution (\ref{solprobe}). The metric $\tilde G_{ij}$ is the following
\begin{equation}
\tilde G_{ij}d\xi^id\xi^j=\frac{\rho^4}{R^4}dx^2_{1,1}+\frac{R^2_{eff}}{\rho^2}d\rho^2
+R_{eff}^2d\Omega_3^2
\label{openmet}
\end{equation}
so it is $AdS_3(R_{eff}/2)\times S^3(R_{eff})$. This is the same metric that was found in \cite{Arean:2007nh} to control the fluctuations of the directions transverse to $S^3$ in $S^7$. As advertised, for $q\neq 0$ it differs from the world-volume induced metric (\ref{wvmetric}), and indeed it resembles the open string metric that controls the kinetic term of the fluctuations of world-volume fields of D-branes.
 
From the fluctuation Lagrangian (\ref{fluclag}), we obtain the equation of motion for $h(\xi)$, and since we are requiring $h(\xi)$ to be independent of the $S^3$ coordinates, this equation of motion boils down to
$$
\partial_\rho(\rho^9\partial_\rho h)+R^4R_{eff}^2\rho^3 (-\partial_0^2+\partial_1^2)h=0
$$
This equation is immediate to solve:  we separate variables and look for a solution of the form $h(x,\rho)=e^{ikx}f(\rho)$. The resulting equation for $f(\rho)$ has two solutions; to pick the relevant one for our problem, we consider their behavior near the boundary, as $\rho \rightarrow \infty$. It turns out that one solution decays as $1/\rho^8$, while the other one becomes independent of $\rho$ as $\rho \rightarrow \infty$, becoming a plane wave at the boundary. This second solution captures variations of the shape of the boundary defect, so it is the one we pick. Explicitly,
\begin{equation}
h(x,\rho)=C e^{ikx}\frac{k^2}{\rho^4}Y_2\left(\frac{R^2R_e|k|}{2\rho^2}\right)
\label{flucsol}
\end{equation}
To make contact with the discussion in the previous section, and the work of \cite{Garriga:2009hy}, note that the fluctuation $h(\xi)$ is not a proper displacement, as seen by an observer in the unperturbed probe brane. To relate it to a proper distance, we compute the unit vector  $n^\mu$ normal to the M5-brane profile (\ref{solprobe}) 
$$
(n^\rho,n^2)=\left(\frac{q\rho}{R^3R_3}, \frac{R^3}{\rho^2 R_e} \right)
$$
so we can introduce a scalar $\phi(\xi)$ measuring proper displacement by defining $h=\phi n^2$ \cite{Garriga:1991ts}. For this scalar field, the equation of motion is
$$
\left(\Box_{AdS_3}-\frac{3}{R^2_{AdS_3}}\right)\phi=0
$$
Notice that now this is a massive scalar field (as the fluctuation now has to climb the AdS well), while for the bubble in deSitter space, the scalar was tachyonic, see eq. (\ref{tacmass}). 
 
\subsection{The regularized action}
Having found the solution for the relevant fluctuation, our next task is to evaluate the action at this solution. As it is by now familiar, an straightforward evaluation yields a divergent result, so we regulate it by introducing a cut-off on $\rho$, $\rho_c$. The regulated action contains pieces analytic in $\rho_c$ that diverge in the limit $\rho\rightarrow \infty$, and a term logarithmic in $\rho_c$, which induces the anomaly in the finite piece. It is the coefficient of this $\log \rho_c$ we are after, since as we will show, it is $d_1/2$. 

The regularized action for the fluctuations is
$$
W[\rho_c]=-T_{M5}\int d^6 \xi \left(\frac{1}{2}\sqrt{g_{S^3}}\frac{\rho^7}{R^4}\tilde G^{ij}\partial _i h\partial _j h\right)
$$
Since we are considering a fluctuation mode $h(\xi)$ independent of $S^3$ coordinates, the integration over $S^3$ is immediate, yielding
$$
W[\rho_c]=-\frac{2\pi^2}{2R^4}T_{M5}\int d^2x \int _0^{\rho_c} d\rho \rho^7 \tilde G^{ij}\partial _i h\partial _j h
$$
Using the explicit form of the metric $\tilde G$, eq. (\ref{openmet}), integration by parts, and the equations of motion, we obtain
$$
W[\rho_c]=-\frac{1}{2}\frac{2\pi^2\rho_c^9}{R^4R^2_{eff}}T_{M5}\int d^2x \; h(\rho_c)\partial _\rho h|_{\rho_c}
$$
We now Fourier transform,  evaluate the action with the solution (\ref{flucsol}) and obtain
$$
W=-\frac{1}{2}\frac{2\pi^2}{R^4R_e^2}T_{M5}\int d^2k |h(k,\rho_c)|^2 \left[ -R^4R^2_{eff}\rho_c^4+\frac{(R^2R_{eff}|k|)^4}{2}\left(\gamma_E+\hbox{log}\frac{R^2R_e|k|^4}{2\rho^2_c}\right)+{\cal O}(1/\rho_c^2)\right]
$$
The first term is analytic in $\rho_c$, and will be taken care of by a corresponding counterterm. The term relevant for us is the one with log $\rho_c^2$ since its coefficient is essentially $d_1$. To establish the precise relation, in the boundary we consider a fluctuation of the shape of the defect, as a plane wave $h(x)=Ae^{ikx}$. One easily sees that for small fluctuations
$$
K_{ij}K^{ij}-\frac{1}{2}K^2\simeq \frac{1}{2}k^4 h^2
$$
Recalling the definition of the trace anomaly (\ref{confan}), we arrive at

\begin{equation}
\frac{d_1}{2} =\frac{\pi^2}{2}R^4R_{eff}^2T_{M5}
\label{resultdd}
\end{equation}
To translate this result into CFT variables, we use
$$
T_{M5}=\frac{1}{(2\pi)^5 \ell_P^6}\hspace{1cm} \frac{R}{\ell_P}=(32\pi^2 N)^{1/6}
$$
where $\ell_P$ is the eleven-dimensional Planck length, and the quantization condition for the world-volume magnetic flux \cite{Camino:2001at}
$$
\frac{q}{\ell_P^3}=4\pi k
$$
where $k$ is the number of M2-branes on one side of the M5 that recombine with it. We finally conclude that
\begin{equation}
d_1=\frac{1}{\pi}\left(N+\frac{1}{2}k^2\right)
\label{dincft}
\end{equation}
Let's pause to contemplate this result. As advertised, for fixed $N$, the minimum value of $d_1$ corresponds to the case when the vacuum density energy is the same on both sides of the defect (i.e. when $k=0$), which is the kind of behavior one would expect, in clear contrast to what happens in de Sitter, see the discussion below eq. (\ref{resultd}). The key point to understand this difference is the relative plus sign in (\ref{dincft}), which in the supergravity side is the relative plus sign in (\ref{radeff}). 

We can also check if the inequality (\ref{puzzle}) is always parametrically satisfied. To do so, we need the jump in the number of degrees of freedom between the two sides of the wall. In the large N limit, it is possible to compute its scaling from supergravity \cite{Klebanov:1996un}, and one obtains
$$
c\sim N^{3/2} \Rightarrow \Delta c \sim (N+k)^{3/2}-N^{3/2}\sim N^{1/2}k
$$
where in the last step we assumed that $k\ll N$. Comparing this with (\ref{dincft}) we see that now the inequality is satisfied, with the particular parametric dependence $k\sim N^{1/2}$ being the only case when both quantities become comparable.

\section{Discussion}
The main goal of this work was to continue the study of possible implications of the spectrum of bubble fluctuations for a potential holographic dual of an eternally inflating multiverse. The first results \cite{Garriga:2009hy} in this direction didn't conform with intuitive expectations, since for instance, they appear to imply that as the difference in vacuum energies on both sides of the bubble grows, the number of degrees of freedom that 'feel' the bubble decreases \footnote{This has a vague resemblance with another counterintuitive feature of deSitter space: for fixed cosmological constant, empty de Sitter has the maximal entropy possible, and adding a black hole decreases rather than increases the total entropy. See \cite{Banks:2006rx} for some ideas on the possible holographic realization of this feature.}.  In this work we have considered a similar setup, but with both de Sitter geometries (bulk and probe world-volume) traded with Anti de Sitter geometries. In this case we see that the coefficient $d_1$ behaves as expected. At the root of the difference lies the fact that the $dS_3$ world-volume in $dS_4$ has curvature radius smaller or equal to that of $dS_4$, while for the $AdS_3$ world-volume in $AdS_4$ just the opposite is true. Therefore, as long as the coefficient $d_1$ is proportional to this thin wall curvature radius, it behaves conforming the expectations of \cite{Garriga:2009hy} for $AdS$, but not for $dS$.

If the hypothetical Euclidean QFT holographically dual to the multiverse does exist, a 2d defect should
give a formula similar to (\ref{dincft}), but with a minus relative sign, so in this case $d_1$ attains its maximum value when both sides of the wall have the same vacuum energy, as required by the arguments after eq. (\ref{resultd}). Where could this relative minus sign come from, in the hypothetical dual theory? To reconcile the intuition that more ambient fields should feel the defect (or more fields live on the defect) as $\Delta \rho_V$ increases  with the result that $d_1$ decreases, we would need degrees of freedom that contributely {\it negatively} to $d_1$. This suggests that these defect degrees of freedom might be non-unitary; we certainly know of non-unitary 2d CFTs with $c<0$, and while $d_1$ is not quite the $2d$ central charge (as it is the coefficient that accompanies the term with extrinsic rather than intrinsic curvature), this suggestion seems worth exploring. Put differently, the authors of  \cite{Garriga:2009hy} speculated that a way out of their puzzles would be that the evaluation of $d_1$  involved unexpected cancellations on the field theory side. A simple example that might illustrate such cancellations is the worldsheet CFT in string theory, which has total vanishing central charge, due to the sum of two theories, one with $c>0$ and one with $c<0$\footnote{I would like to thank Micha Berkooz for suggesting this example, and for related conversations.} .

At a technical level, the main result in this paper is the computation of the coefficient $d_1$ in the conformal anomaly for a surface operator in a 2+1 CFT, following the prescription of \cite{Garriga:2009hy}. Surface operators in 3+1 dimensional theories have received a lot of attention, but there are very few explicit computations of their parameters \cite{Drukker:2008wr}. The kind of computation performed here might find applications in the study of surface operators, beyond our original motivation.

\section{Acknowledgements} 
This work grew out of many conversations with Jaume Garriga, and I would like to thank him for
many explanations about  \cite{Garriga:2008ks, Garriga:2009hy}. I would also like to thank Ofer Aharony and Micha Berkooz for insightful discussions and Alfonso Ramallo for useful correspondence regarding \cite{Arean:2007nh}. I would like to thank the Department of Particle Physics at the Weizmann Institute of Science for hospitality during the last stages of this project. This research is supported by a Ram\'{o}n y Cajal fellowship, and also by MEC FPA2007-66665C02-02, CPAN CSD2007-00042, within the Consolider-Ingenio2010 program, and AGAUR 2009SGR00168.

\end{document}